\documentclass[doublecol]{epl2}

\title{Light-induced optical switching in an asymmetric metal-dielectric microcavity with phase-change material}
\shorttitle{Light-induced optical switching in a microcavity with phase-change material} 

\author{R. Thomas\inst{1} \and A. A. Chabanov\inst{2} \and I. Vitebskiy\inst{3} \and T. Kottos\inst{1}}
\shortauthor{R. Thomas \etal}

\institute{                    
  \inst{1} Wave Transport in Complex Systems Lab, Department of Physics, Wesleyan University, Middletown, CT 06457, USA\\
  \inst{2} Department of Physics and Astronomy, University of Texas at San Antonio, San Antonio, TX 78249, USA\\
  \inst{3} Air Force Research Laboratory, Sensors Directorate, Wright-Patterson Air Force Base, OH 45433, USA
}

\abstract{
We propose an infrared power switch based on an asymmetric high-Q microcavity incorporating a metallic nanolayer in close proximity 
to a layer made of a phase-change material (PCM). The microcavity is designed so that when the PCM layer is in the low-temperature 
phase, the metallic nanolayer coincides with a nodal plane of the resonant electric field component, to allow a high resonant transmittance. 
As the light intensity exceeds a certain threshold, light-induced heating of the PCM layer triggers the phase transition accompanied by an 
abrupt change in its refractive index in the vicinity of the transition temperature. The latter results in a shift of the nodal plane away from 
the metallic nanolayer, rendering the entire microcavity highly reflective over a broad frequency range. The nearly binary nature of the 
PCM refractive index allows for the low-intensity resonant transmission over a broad range of ambient temperatures below the transition point.}

\begin{document}

\maketitle

\section{Introduction}
Degeneracies --€" both symmetry related and accidental --€" have been utilized in a variety of physical settings. Examples 
include the use of: (i) diabolic points for dispersion-less propagation and relativistic type of transport \cite{BJ07,PBFMSC07,TPGSSB10}, 
(ii) exceptional points of degeneracy for hypersensitive sensors via spontaneous PT-symmetry violation \cite{FGG17,GMKMRC18}, and 
(iii) accidental spectral degeneracies for the design of channel drop filters \cite{FVJH98a,FVJH98b} and perfect absorbers \cite{PLF14,PF14}. 

In this Letter, we employ accidental degeneracy and degeneracy breaking to realize a new class of photonic switches. In contrast to the above-mentioned 
examples, we use accidental degeneracy in the spatial domain. Specifically, we exploit the coincidence of a thin metallic nanolayer 
with a (quasi-)nodal plane of a resonant mode supported by a microcavity. Importantly, the metallic nanolayer location is asymmetric, which 
makes its coincidence with the nodal plane of the resonant field distribution accidental. Due to the presence of a phase-change material 
(PCM) layer in close proximity to the metallic nanolayer, the spatial accidental degeneracy (SAD) only occurs at low 
incident light intensity, when the PCM is in the low-temperature phase. If the incident light intensity exceeds a certain threshold, the light-induced heating 
causes the transition of the PCM to the high-temperature phase, with an abruptly changing refractive index. As a consequence, the nodal plane 
of the resonant field distribution is shifted away from the metallic nanolayer, lifting the SAD and completely suppressing the resonant mode along 
with the resonant transmission. As a result, the microcavity turns highly reflective. It is essential that we use the PCM in the microcavity design to switch 
from the low-intensity resonant transmission to the high-intensity broadband reflection. Indeed, if instead of the PCM we used a material 
with a temperature-dependent refraction (as was the case in Ref. \cite{MCVK16}), then the SAD and the related low-intensity resonant 
transmission could be realized only at a particular temperature. By comparison, the nearly binary nature of some practically available 
PCMs, such as vanadium oxides, allows for the low-intensity resonant transmission over a relatively broad range of ambient temperatures.

\section{Photonic crystal design}
\begin{figure}
\onefigure[width=8.5cm]{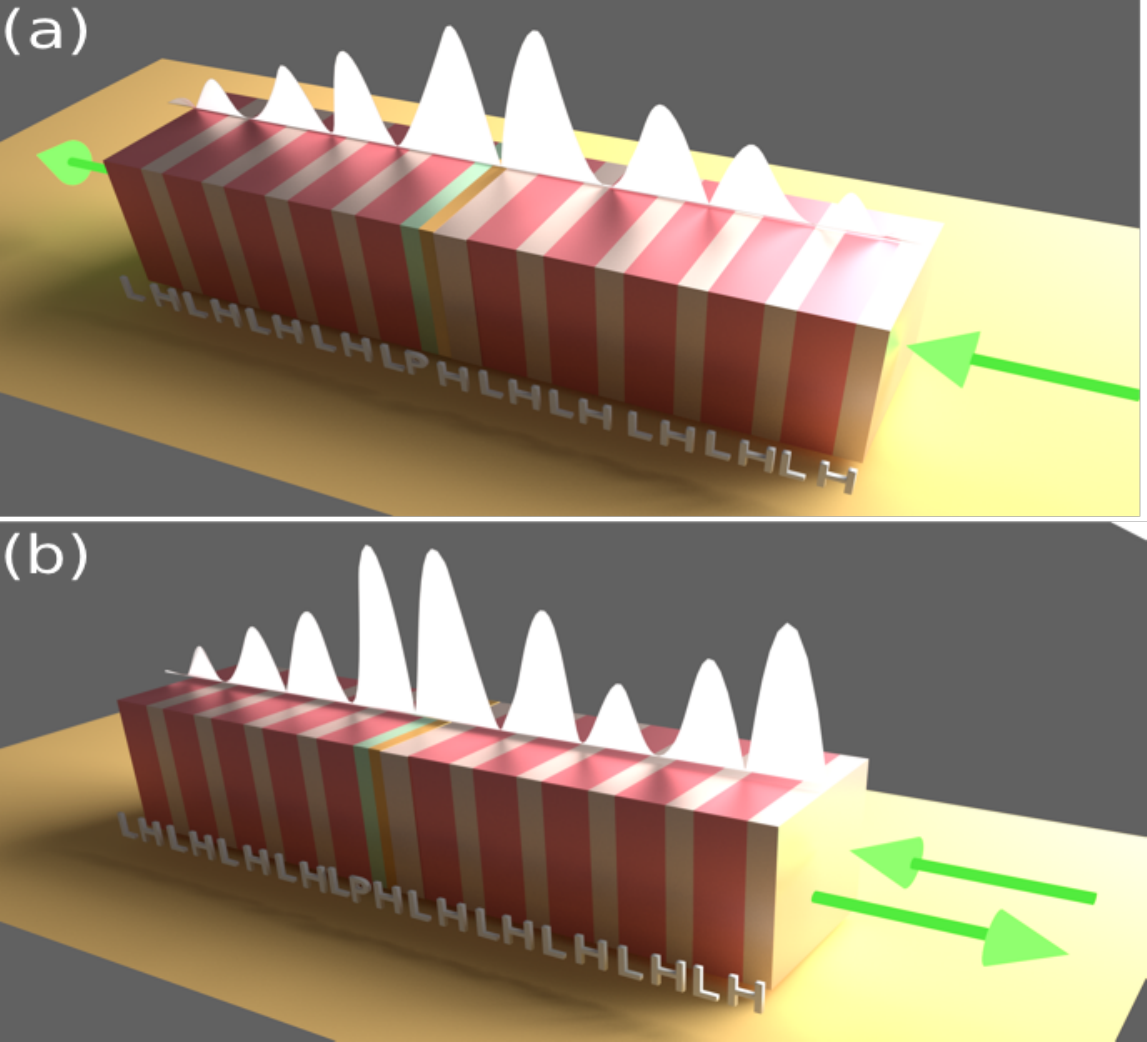}
\caption{Schematics and operating principle of the photonic switch based on asymmetric metal-dielectric planar microcavity with the phase-change material, 
VO$_2$. The microcavity consists of quarter-wave layers of high ($H$, light orange) and low ($L$, red) refractive index materials and incorporates 
a metallic nanolayer in close proximity to the VO$_2$ layer ($P$, greenish). (a) When the VO$_2$ layer is in the low-temperature phase, the metallic nanolayer 
coincides with a nodal plane of the resonant electric field component (shown white), to allow a high resonant transmittance. (b) When the VO$_2$ is in the 
high-temperature phase, the nodal plane shifts away from the metallic nanolayer, resulting in a broadband reflection of the photonic switch. The transition 
from the low- to the high-temperature phase of the VO$_2$ layer is caused by the light-induced heating. The spatial intensity distributions of the electric 
field component are shown using a logarithmic scale.}
\label{fig1}
\end{figure}

The proposed photonic switch is shown in Fig. \ref{fig1} and it is designed to operate at a mid-infrared (MIR) wavelength of 10.6 $\mu$m, 
at which the highest-power continuous-wave (CW) CO$_2$ lasers are available. It consists of quarter-wave layers of Si ($H$, light orange) 
and ZnS ($L$, red) of the respective refractive indices, $n_{\rm Si}\approx 3.42$\cite{Si_index} and  $n_{\rm ZnS}\approx 2.20$\cite{ZnS_index}, 
and thicknesses, $d_{\rm Si}$=775 nm and  $d_{\rm ZnS}$=1.20 $\mu$m. Their respective extinction coefficients are negligibly small in the MIR. 
An asymmetric defect/cavity is introduced in the middle of the photonic crystal. It consists of a 1.70-$\mu$m-thick Si layer and a 1.95-$\mu$m-thick 
layer of vanadium dioxide (VO$_2$) ($P$, greenish), which undergoes a phase transition 
from the insulating to the metallic phase at a critical temperature of $\theta_{\rm c}\approx 341$ K. At $\theta_{\rm c}$, the real and imaginary parts of 
the refractive index of VO$_2$, $n_{\rm VO_2}^{\prime}$ and $n_{\rm VO_2}^{\prime\prime}$ respectively, exhibit large changes, as determined from the 
MIR experimental data of Ref. \cite{K13}. The design of the photonic switch is completed with a 5-nm thick metallic layer, strategically incorporated in 
the asymmetric microcavity. The metallic layer is placed in such a way that below $\theta_{\rm c}$, it coincides with one of the (quasi-)nodal planes 
of the electric field component of a defect-localized mode supported by the microcavity at 10.6 $\mu$m. In the numerical example, we assume that 
the metallic nanolayer is made of copper (Cu) with the permittivity $\epsilon_{\rm{Cu}}=(-4.26+i1.19)\times 10^3$ at 10.6 $\mu$m \cite{OLBBBAW83,OBALQ}.

\section{Maxwell's-Heat Transfer modeling}
The electromagnetic wave propagation in a photonic structure with temperature-dependent 
constituent parameters is obtained from the following coupled differential equations \cite{TVK17}:
\begin{eqnarray}
\boldsymbol{\nabla}\times\boldsymbol{H} = {\boldsymbol j}_0+ \epsilon(z,\theta)\frac{\partial\boldsymbol{E}}{\partial t},\quad
\boldsymbol{\nabla}\times\boldsymbol{E} = -\mu_0\frac{\partial{\boldsymbol{H}}}{\partial t}, \label{maxwell}\\
\rho_D C_{p}^D\frac{\partial\theta}{\partial t} -\boldsymbol{\nabla}\cdot(k_D\boldsymbol{\nabla}\theta) =Q +q_0+q_r\nonumber,
\end{eqnarray}
where $\epsilon(z,\theta)=\epsilon^{\prime}(z,\theta)+i\epsilon^{\prime\prime}(z,\theta)$ describes the temperature-dependent
permittivity along the propagation direction $z$, $\mu_0$ is the permeability of free space, and ${\boldsymbol j}_0=\sigma(z){\boldsymbol E}$ is 
the electric current. The quantity $\epsilon^{\prime\prime}(z,\theta)$ determines non-ohmic losses. 
$C_D^{P}= 700$ J/(kg$\cdot$K), $\rho_D=4571$ kg/m$^3$ and $k_D=4$ W/(m$\cdot$K) are the specific heat capacity, 
mass density and thermal conductivity, respectively, in the insulating phase of VO$_2$ \cite{CGZZ11,CE15,LRVFL07}. 
The self-induced heating via the incident radiation leads to the heat production, 
$Q= \frac{1}{2}\left({\cal R}e(\boldsymbol{j}_0\!\cdot\boldsymbol{E})+\omega\epsilon^{\prime\prime}(z)|\boldsymbol{E}|^2\right)$. 

In our simulations, we also take into account the convective heat flux, $q_0=h(\theta_{\rm ext}-\theta)$, from the outer layers of the photonic 
structure to the surrounding air, which is assumed to be at the ambient temperature, $\theta_{\rm ext}=293$ K. The heat flux coefficient is 
$h=10$ W/(m$^2\cdot\rm{K}^4$) \cite{wiki}. Finally, the term $q_r= \epsilon_r\sigma_r(\theta_{\rm ext}^4 - \theta^4)$ describes the heat transfer via thermal radiation from the 
outer layers to the surrounding air, assuming gray-body approximation \cite{LVSB13}. The parameters $\epsilon_r$ and $\sigma_r = 
5.67 \times 10^{-8}$ W/(m$^2\cdot\rm{K}^4$) correspond to the thermal emissivity coefficient of the outer layers and the Stefan-Boltzmann 
constant, respectively. The thermal emissivity $\epsilon_r$ of the outer (Si and ZnS) layers was estimated from Kirchhoff's 
law and was considered using a self-consistent evaluation scheme \cite{ATKVK18}. 

Eqs. (\ref{maxwell}) are solved numerically, at a given wavelength, using a finite-element software package from COMSOL 
MULTIPHYSICS \cite{comsol}, in order to evaluate the time-dependent transmittance $T(t)$, reflectance $R(t)$ and absorptance 
$A(t) = 1 - T(t) - R(t)$. The corresponding asymptotic (steady-state) values, $T_{\infty}$, $R_{\infty}$ and $A_{\infty}$, have been 
extracted from the temporal dynamics after ensuring that the steady state has been reached. These steady-state values have been found 
to agree well with $T_{\infty}$, $R_{\infty}$ and $A_{\infty}$, calculated using an alternative approach based on a Frequency-Stationary 
module \cite{comsol} coupled to heat transfer equations. 

In the simulations we used a varying mesh density: the Si and ZnS layers  have been partitioned with 50-150 elements per layer, depending on the simulation, 
while the cavity defect layers have been partitioned with 300-1000 elements. The convergence of the results has been evaluated with a tolerance factor which has been set to 0.1\%. 
We have furthermore repeated the calculations by doubling the number of mesh points in order to guarantee the accuracy of the converged numerical solutions.

\section{Results and analysis}
\begin{figure}
\onefigure[width=8.8cm]{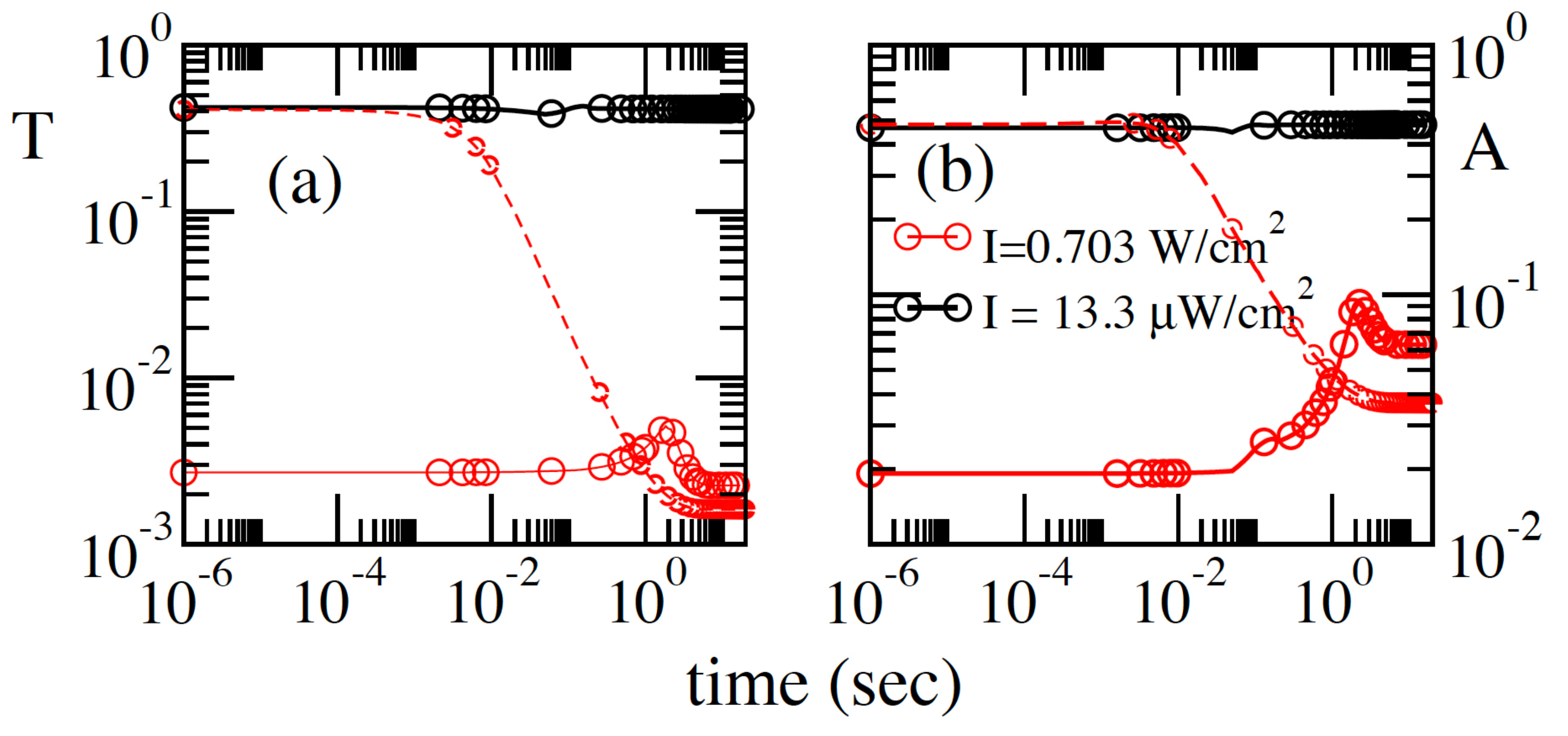}
\caption{Temporal dynamics of transmittance $T(t)$ and absorptance $A(t)$ of incident radiation of $I=13.3$ $\mu$W/cm$^2$ (black line-circles) and $I=0.703$ W/cm$^2$ 
(red dashed-line-circles) at the microcavity wavelength 10.6 $\mu$m. The red solid-line-circles correspond to $T(t)$ and $A(t)$ of the high-intensity incident wave at 
the resonance frequency of the microcavity with the asymptotic (steady-state) value of the refractive index of VO$_2$, ${n_{\rm VO_2}^{\prime}}_\infty$.}
\label{fig2}
\end{figure}

In Fig. \ref{fig2}, we show two examples of the temporal dynamics of transmittance $T(t)$ and absorptance $A(t)$ at the microcavity wavelength 10.6 $\mu$m. 
The first example corresponds to low incident intensity of $I=13.3$ $\mu$W/cm$^2$ (black line-circles), the second to high incident intensity of 
$I=0.703$ W/cm$^2$ (dashed red line-circles). In the low-intensity case, both the transmittance and absorptance do not change noticeably with time, 
indicating that almost no variation in $n_{\rm VO_2}^{\prime}$ occurs due to thermal effects. In the high-intensity case, the short-time transmittance is of $\sim{\cal O}(1)$, 
while at longer times it acquires values of $\sim{\cal O}(10^{-3})$ due to the dynamical detuning associated with the thermal variation of $n_{\rm VO_2}^{\prime}$.
In the steady state, the resonance frequency is determined by the asymptotic value of the refractive index of VO$_2$, ${n_{\rm VO_2}^{\prime}}_\infty$. At this frequency, 
the transmittance (red solid-line-circles) remains $\sim {\cal O}(10^{-3})$ at all times, as at short times the incident wave is detuned from the microcavity resonance,
while at longer times a transition to the over-damping regime (see below) dominates the transport characteristics, leading to suppression of any resonant effect. 
The temporal behavior of the absorptance is qualitatively similar to that of the transmittance. It is worth noting that in Fig. \ref{fig2}(a) the activation time of switching 
(i.e., time at which transmittance drops down from a near-unity value) is $\sim 10^{-2}$ s. One can further decrease the activation time of switching by increasing 
the number of layers of the Bragg reflectors. In fact, a recent analysis indicates that the activation time scales exponentially with the number of layers \cite{ATKVK18}.

\begin{figure}
\onefigure[width=8.8cm]{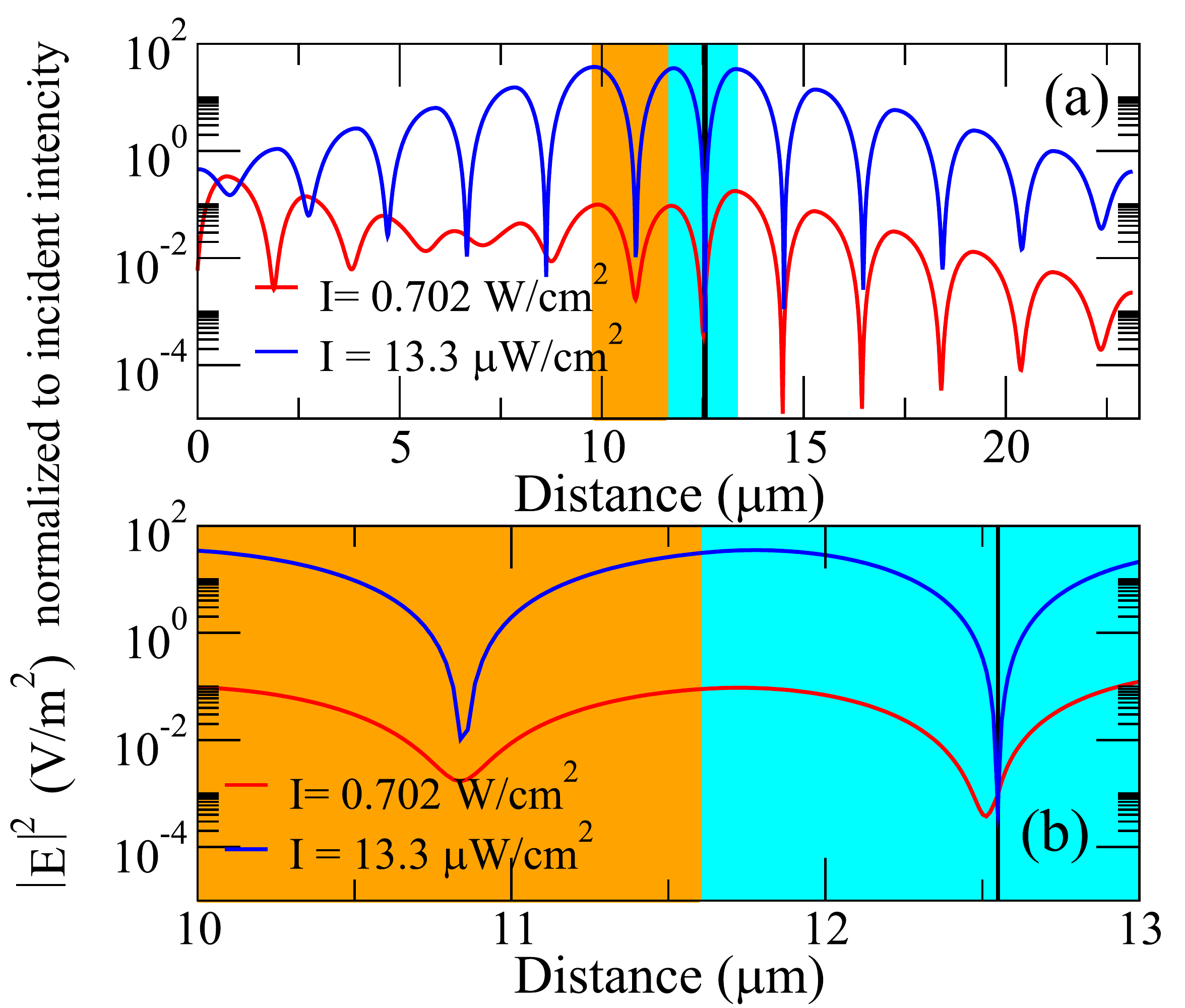}
\caption{(a) Spatial intensity distribution of the resonant electric field component for the case of low and high incident intensities, $I=13.3$ $\mu$W/cm$^2$ (blue line) 
and $I=0.702$ W/cm$^2$ (red line) respectively, at the microcavity wavelength 10.6 $\mu$m. Only the central part of the photonic switch is shown 
(the VO$_2$ layer is orange, Si layer is cyan, and the metallic nanolayer is the black line). (b) The zoomed-in intensity distributions of the resonant electric field.}
\label{fig3}
\end{figure}

To get a better understanding of the different behavior of the photonic switch at low and high incident intensities, we now turn to the study of 
the scattered field. In Fig. \ref{fig3}a, we show the respective intensity distributions of the resonant electric field inside the photonic switch. For 
simplicity, we highlight only the region of the photonic structure which is associated with the defect cavity (the VO$_2$ layer is shown in orange, 
the Si layer in cyan, and the metallic nanolayer is indicated by the black line). In the case of low incident intensity ($I=13.3$ $\mu$W/cm$^2$), the scattered field 
(Fig. \ref{fig3}a, blue line) is localized at and decays exponentially away from the cavity. The localized mode has a long but finite lifetime and 
develops (quasi-)nodal planes inside the cavity. At the nodal planes, the electric field intensity is essentially approaching zero. By positioning the metallic nanolayer 
at the nodal-plane position, we induce SAD. A straightforward calculation allows us to evaluate light absorption being proportional to 
$ \int dz |E(z)|^2 \epsilon^{\prime\prime}(z)$. Clearly, in the case of SAD, light absorption is minimal. Consequently, transmittance is high and occurs 
via the resonant localized mode (with a Q factor of $\sim 500$), which is not significantly affected by the presence of the metallic layer.

In Fig. \ref{fig3}a, we also show the intensity distribution of the steady-state scattered field (red line) for the high incident intensity ($I=0.702$ W/cm$^2$)
at the same resonance wavelength 10.6 $\mu$m. In this case, the field decays exponentially from the incident surface of the photonic switch, indicating that 
the localized mode is completely suppressed. Specifically, we find that the SAD has been lifted due to a shift of the nodal plane away from the 
metallic nanolayer (see the zoomed-in spatial intensity distribution in Fig. \ref{fig3}b). Consequently, the electric field interacts with the metallic nanolayer, 
leading to a near-unity reflectance and vanishing transmittance.

\begin{figure}
\onefigure[width=8.8cm]{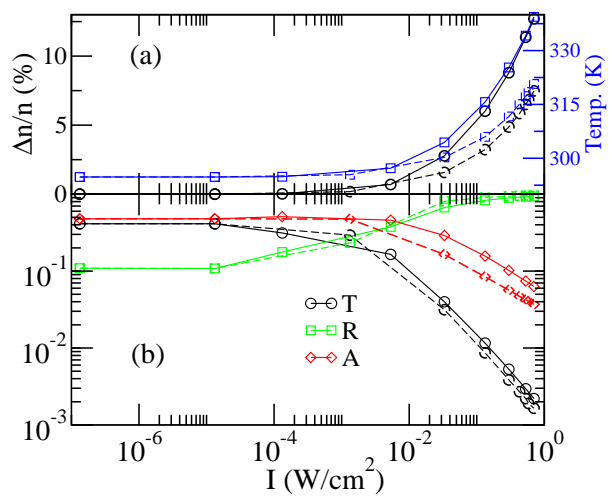}
\caption{(a) The absolute temperature of the VO$_2$ layer (blue lines) and the percentage variation of the refractive index ${n_{\rm VO_2}^{\prime}}_\infty$ 
(black lines) as a function of the incident light intensity $I$. Notice that VO$_2$ temperature is below the transition point $\theta_{\rm c}\approx 341$ K. 
(b) The steady-state transmittance ${T}_{\infty}$ (black line-circles), reflectance ${R}_{\infty}$ (green line-circles) and absorptance ${A}_{\infty}$ (red line-circles) 
as a function of $I$. Notice steep decrease of ${T}_{\infty}$ and ${A}_{\infty}$ and increase in ${R}_{\infty} $ for $I > 10^{-4}$ W/cm$^2$. The dashed and 
solid lines correspond to the incident light intensity at the resonance frequency of the unperturbed and the most-perturbed cavity, respectively.}
\label{fig4}
\end{figure}

The physical mechanism that triggers the shift of the nodal plane and the lifting of the SAD is understood once we analyze the behavior of 
${n_{\rm VO_2}^{\prime}}_\infty$ as a function of the incident light intensity.  A direct numerical evaluation, using Eqs. (\ref{maxwell}), of the temperature 
of the VO$_2$ layer (Fig. \ref{fig4}a, blue line-squares) and the resulting variation of ${n_{\rm VO_2}^{\prime}}_\infty$ (Fig. \ref{fig4}a, black line-circles) 
at the resonance frequencies of the unperturbed and most-perturbed microcavities (dashed and solid lines, respectively) clearly indicate that 
$n_{\rm VO_2}^{\prime}$ has been modified due to resonance-enhanced light-matter interaction, altering the optical properties of the microcavity 
and the position of the nodal planes of the scattered field. 

The steady-state transport characteristics of the photonic switch as a function of the incident light intensity are reported up to the transition 
temperature $\theta_c$ in Fig. \ref{fig4}b. For CWs of low intensity, $I< 10^{-4}$ W/cm$^2$, the photonic switch exhibits high transmittance $T_{\infty}$ at the 
resonant cavity frequency (Fig. \ref{fig4}b, black, solid and dashed line-circles). However, as the light intensity is increased up to $\sim 10^{-2}$ W/cm$^2$, 
the self-induced heating causes a variation of ${n_{\rm VO_2}^{\prime}}_\infty$ (see Fig. \ref{fig4}a), which is large enough to lift the SAD and suppress the 
resonant  cavity mode. As the incident light intensity is further increased, the system undergoes a transition from under-damping to over-damping. 
The resulting impedance mismatch between the incident field and the resonant cavity mode affects dramatically the transport characteristics of the switch. 
For incident light intensities $I>10^{-2}$ W/cm$^2$, there is a steep decrease in the steady-state transmittance (Fig. \ref{fig4}b, black, solid and 
dashed line-circles) and absorptance (Fig. \ref{fig4}b, red, solid and dashed line-diamonds), while the reflectance acquires near-unity values 
(Fig. \ref{fig4}b, green, solid and dashed line-squares). Also notice in Fig. \ref{fig4} that the lifting of the SAD requires a very moderate increase in the 
incident light intensity, thus making a very sensitive optical switch. On the other hand, the nearly binary nature of the refractive index of VO$_2$ allows 
for the low-intensity resonant transmission over a relatively broad range of ambient temperatures below the transition point.

\section{Conclusions}
We have proposed a photonic limiter-switch based on asymmetric metal-dielectric planar microcavity with a PCM component (VO$_2$ in our case). 
The cavity is designed so that at low incident light intensity, when the VO$_2$ layer is in the low-temperature phase, the location of the metallic nanolayer 
coincides with the nodal plane of the resonant electric field distribution, thus allowing for high resonant transmittance. As the incident light intensity increases, 
the VO$_2$ layer approaches the phase transition temperature, and its refractive index changes significantly -- by up to 10\%, even before reaching the transition 
point (see Fig. 4a). As a consequence, the nodal plane of the electric field distribution shifts away from the metallic nanolayer, rendering the planar cavity highly reflective 
within a broad frequency range. The use of the PCM, such as VO$_2$, rather than an optical material with gradual temperature dependence of the refractive index, 
is critically important because it provides a broad temperature range below the transition temperature $\theta_c$, where the refractive index is virtually temperature-independent (see Fig. 4a). 
This circumstance allows to maintain the SAD condition, along with the strong resonant transmission, within the broad temperature range. By comparison, in the 
case of a gradual temperature dependence of the refractive index \cite{MCVK16}, the SAD conditions and the corresponding resonant transmittance of the metal-dielectric 
cavity can be achieved only at a certain temperature. Yet another possibility is to replace the PCM with an optical material displaying strong Kerr nonlinearity. 
This and similar approaches (see, for example, \cite{LU11} and references therein) could be more appropriate in the case of short high-intensity pules, where the effect 
of heating on the refractive index is less significant than the Kerr nonlinearity or nonlinear absorption. Finally, instead of light-induced change in the refractive index, 
one could use materials with nonlinear or temperature dependent absorption (see, for example, \cite{LIU11} and references therein). These designs, though, cannot take 
advantage of the sensitivity-enhancing metallic nanolayers, because nonlinear absorption is much less effective in the lifting of SAD, compared to the temperature 
related change in the refractive index. In addition, the presence of lossy components can result in the device overheating and damage.

Due to the use of metallic nanolayer under the SAD conditions, the proposed cavity design shows sharp transition from the low-intensity resonant transmittance 
to high-intensity reflectivity even before the VO$_2$ layer reaches the transition temperature $\theta_c$. If the input light intensity is further increased so that the VO$_2$ layer 
completely transitions to the high-temperature phase, the expected jump in the cavity reflectivity will be even higher. We can also expect additional effects, such 
as optical bistabilities and strong asymmetry between the cases of increasing and decreasing input light intensity. These effects, though, will be addressed 
in a separate publication. The main objective of the present study is to demonstrate the effect of SAD associated with the metallic nanolayer judiciously positioned 
in a planar microcavity with a PCM component.

\acknowledgments
This research was partially supported by DARPA NLM program via grant No. HR00111820042 and by an ONR 
N00014-16-1-2803 (R. T., A.C. \& T. K.). (A.C.) also acknowledges partial support from AFOSR via FA9550-16-1-0058 grant. (I.V.) was 
supported by an AFOSR LRIR 18RYCOR013 grant.

\end{document}